\documentclass[a4paper,twocolumn,prb]{revtex4}
\usepackage{graphicx}
\usepackage{xspace}
\begin{document}
\title{Large two-level magnetoresistance effect in doped manganite
grain boundary junctions}
\author{J.~B.~Philipp$^{1,2}$, C.~H\"{o}fener$^1$, S.~Thienhaus$^1$, J.~Klein$^1$,
L.~Alff$^{1,2}$, and R.~Gross$^{1,2}$}
\affiliation{$^1$II.~Physikalisches Institut, Universit\"{a}t zu K\"{o}ln,
Z\"{u}lpicher Str.~77, D - 50937 K\"{o}ln, Germany}

\affiliation{$^2$Walther-Mei{\ss}ner-Institut f\"{u}r Tieftemperaturforschung,
                   Walther-Mei{\ss}ner Str.~8, D - 85748 Garching, Germany}

\date{received \today}


\begin{abstract}
We performed a systematic analysis of the tunneling
magnetoresistance (TMR) effect in single grain boundary junctions
formed in epitaxial La$_{2/3}$Ca$_{1/3}$MnO$_3$ films deposited
on SrTiO$_3$ bicrystals. For magnetic fields $H$ applied parallel
to the grain boundary barrier, an ideal two-level resistance
switching behavior with sharp transitions is observed with a TMR
effect of up to 300\% at 4.2\,K and still above 100\% at 77\,K.
Varying the angle between $H$ and the grain boundary results in
differently shaped resistance vs $H$ curves. The observed
behavior is explained within a model of magnetic domain pinning
at the grain boundary interface.

\end{abstract}

\maketitle

Ferromagnetic tunneling junctions have been studied intensively over the
last years due to possible applications in magnetoelectronics devices. The
tunneling magnetoresistance (TMR) between two ferromagnetic layers $i=1,2$
separated by a thin insulating barrier depends on the relative orientation
of the magnetization and the spin polarization $P_i=2a_i-1$, where $a_i$ is
the fraction of majority spin electrons in the density of states of layer
$i$. Within the Julli\`{e}re model \cite{Julliere:75a}, the TMR is estimated to

\begin{equation}
\displaystyle\frac{\Delta
R}{R}=\displaystyle\frac{R_{\uparrow\downarrow}
-R_{\uparrow\uparrow}}{R_{\uparrow\uparrow}}=
\displaystyle\frac{2P_1P_2}{1-P_1P_2}\; ,
 \label{Julli}
\end{equation}

\noindent where $R_{\uparrow\uparrow}$ and $R_{\uparrow\downarrow}$ is the
tunneling resistance for parallel and anti-parallel magnetization
orientation. Depending on the spin polarization, for parallel magnetization
the tunneling resistance is significantly reduced, since the large density
of occupied and empty states for the majority spin electrons in both
junction electrodes allow for a large tunneling current. We note that in the
Julli\`{e}re model only elastic tunneling without any spin-flip processes is
assumed and the junction electrodes are assumed to be single domain.

So far, most of the investigated TMR devices are based on
transition metals and compounds such as Ni, Co, Fe, or
Co$_{50}$Fe$_{50}$ with $P\lesssim 50$\% (for recent overviews
see Moodera {\em et al.} \cite{Moodera:99} and Parkin {\em et
al.} \cite{Parkin:99a}). It is evident that $P$ close to unity is
desired to achieve a high TMR-effect. There are several candidates
for materials with large $P$ close to 100\% such as the Mn-based
Heusler alloys \cite{deGroot:83}, the oxide ferromagnets as
Fe$_3$O$_4$ or CrO$_2$, and the doped manganites of composition
La$_{x}D_{1-x}$MnO$_3$ with $D=$ Ca, Sr, and Ba.  While for the
former materials the high spin polarization is still under
question \cite{Kaemper:88}, recently photoemission spectroscopy
has provided direct evidence for the half-metallic nature of
La$_{0.7}$Sr$_{0.3}$MnO$_3$ \cite{Park:98} with $P$ close to
unity. Indeed, using doped manganites high TMR values have been
achieved \cite{Lu:96,Sun:96}, including trilayer spin valve
devices with a TMR value above 450\% at 4.2\,K corresponding to a
spin polarization above 80\% \cite{Viret:97a,Blamire:2000} and
even above 1000\% at 4.2\,K corresponding to a spin polarization
above 90\% \cite{Lu:00}.  We emphasize that the TMR-effect in
manganite tunnel junctions is observed at low applied magnetic
field of less than 100\,mT and has to be distinguished from the
intrinsic, high-field colossal magnetoresistance (CMR) of the
doped manganites \cite{vonHelmolt:93,Jin:94}. While most tunnel
junctions rely on multilayer technology, one can also form
ferromagnetic tunnel junctions by using well-defined individual
grain boundaries (GBs) separating two ferromagnetic grains. This
is achieved by growing epitaxial manganite films on a SrTiO$_3$
bicrystal substrates
\cite{Mathur:97a,Steenbeck:97,Klein:99a,Gross:99a,Hoefener:00}.
In this configuration the barrier is formed by a straight, few nm
wide distorted GB interface as shown by transmission electron
microscopy\cite{Gross:99a,Wiedenhorst:00,Gross:00a}. After
annealing in oxygen atmosphere, single GB junctions (GBJs) with
large TMR effects have been
achieved\cite{Mathur:97a,Steenbeck:97,Klein:99a,Gross:99a,Hoefener:00}.

For many applications not only a large TMR effect but also a sharp switching
between two distinct resistance values at a well-defined magnetic field is
required.  Such ideal, almost rectangular shaped $R(H)$-curves with a
maximum resistance change below 50\% \cite{Lu:96} have been observed for
some devices based on transition metals. However, for tunnel junctions based
on doped manganite junctions (and also for many devices based on transition
metals), usually strongly rounded and noisy $R(H)$ curves \cite{Sun:96},
sharply peaked structures \cite{Mathur:97a}, or multiple resistance level
switching is reported. This behavior most likely originates from an
uncontrolled domain switching. Here, we present a detailed study of the
shape of the $R(H)$ curves of GBJs based on doped manganites. We show that
an almost perfect two-level resistance switching behavior with a sharp
transition between the resistance values can be obtained. We further show
that such ideal behavior is achieved with the magnetic field applied within
the film plane parallel to the GB barrier. We also demonstrate that it is
possible to vary the magnitude of the switching field $H_s$ by changing the
angle between the GB and the applied field. Our results provide evidence
that domain wall pinning structures, intergrain coupling, and the direction
of the applied field can be combined to taylor the $R(H)$-curves of magnetic
tunnel junctions.

GBJs were fabricated by pulsed laser deposition of epitaxial, 80\,nm thick
La$_{2/3}$Ca$_{1/3}$MnO$_3$ films on symmetrical, [001] tilt SrTiO$_3$
bicrystals with a misorientation angle of 24$^{\circ}$. After film
deposition the samples were annealed ex-situ at 950$^{\circ}$C in oxygen
atmosphere for one hour. X-ray analysis of the films shows only (00$\ell$)
reflexes and the FWHM of the (002) rocking curve was 0.04$^{\circ}$.
Microbridges of 30\,$\mu$m width straddling the GB were patterned using
optical lithography and Ar ion beam etching. In this way, well-defined
individual GBJs were obtained. The La$_{2/3}$Ca$_{1/3}$MnO$_3$ films
typically had a Curie-temperature $T_C$ of about 225\,K. Further details on
the transport properties and the microstructure of the GBJs have been
reported recently
\cite{Klein:99a,Gross:99a,Hoefener:00,Wiedenhorst:00,Gross:00a,Wiedenhorst:99a}.

\begin{figure}[h]
\centering{%
\includegraphics [width=0.8\columnwidth,clip]{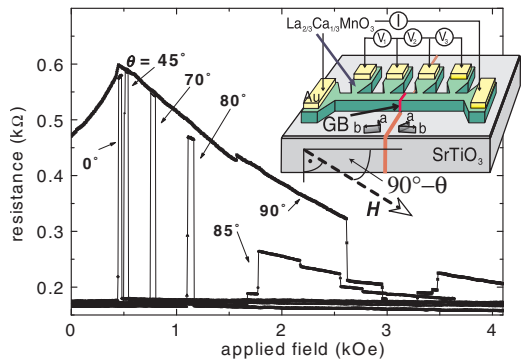}}
 \vspace*{-3mm}\\
 \caption{
$R(H)$ curves of a La$_{2/3}$Ca$_{1/3}$MnO$_3$ GBJ for different
angles $\theta$ between $H$ and the GB barrier measured at
$4.2$\,K and a bias voltage of 5\,mV. For clarity, the $R(H)$
curve is shown only for one direction of the field sweep. The
inset show a sketch of the GBJ geometry.
 }
 \label{angle}
\end{figure}

In Fig.~\ref{angle}, typical $R(H)$ curves of a
La$_{2/3}$Ca$_{1/3}$MnO$_3$ GBJ are shown for different angles
$\theta$ between the GB barrier and the magnetic field $H$, which
always is applied within the film plane. For
$0^\circ<\theta<80^\circ$ rectangular shaped $R(H)$-curves are
observed clearly indicating a two-level resistance switching
behavior. Here, $\theta =90^\circ$ and $0^\circ$ corresponds to
$H\perp {\rm GB}$ and $H\| {\rm GB}$, respectively. The switching
field $H_{s}$ is shifted to larger values with increasing
$\theta$. At the same time the TMR effect decreases. For $\theta
\gtrsim 80^\circ$ several sharp jumps occur in $R(H)$ indicating a
multi-level resistance scheme. At $\theta=90^{\circ}$, where
$H\perp {\rm GB}$, a broad continuous $R(H)$ curve is observed
forming the envelope of the $R(H)$ curves measured for $\theta <
90^\circ$. In the $\theta=90^{\circ}$ curve still some jump-like
resistance switches between well-defined states are observed.  The
field scale for $H_s$ is set by the coercivity field $H_c$ which
varies for differently doped manganites. In addition, the measured
switching field $H_s$ depends on the velocity of the field sweep
and the magnetic history indicating the importance of domain
dynamics.

\begin{figure}[h]
\centering{%
\includegraphics [width=0.80\columnwidth,clip]{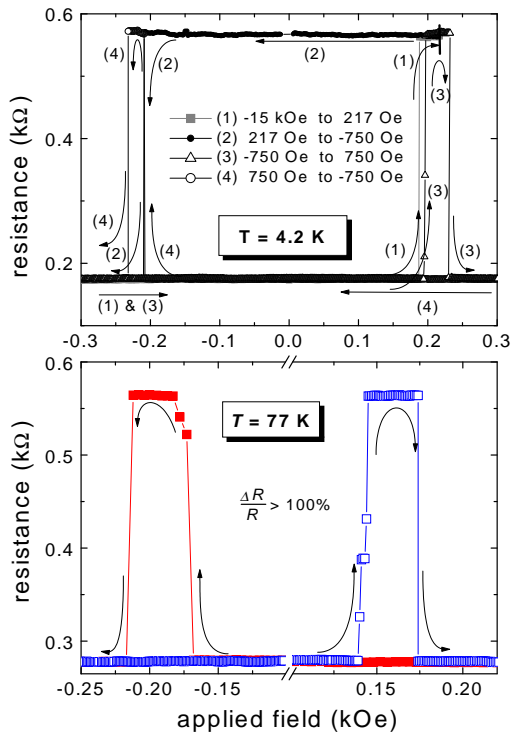}}
 \vspace*{-3mm}\\
 \caption{
$R(H)$-curves for $\theta=0^{\circ}$ at 4.2\,K (upper panel) and
77\,K (lower panel). The direction of the field sweep is indicated
by the arrows.
 }
 \label{large}
\end{figure}

According to Julli\`{e}re's model \cite{Julliere:75a}, a two-level
behavior of the $R(H)$-curves is expected if the electrodes are
considered to be single domain. The two resistance levels
correspond to the fully parallel and anti-parallel magnetization
configuration of the two ferromagnetic electrodes. That is, the
$R(H)$-curves observed for $\theta=0^\circ$ are close to those
expected according to the Julli\`{e}re model for a tunnel junction
consisting of electrodes with slightly different $H_c$. The $R(H)$
curves for $\theta=0^{\circ}$ are shown in more detail in
Fig.~\ref{large} for several field sweeps. The switching between
the two resistance levels occurs within less than 10\,Oe at
4.2\,K.  When the direction of the field sweep is reversed on the
high-resistance branch (curve 1), the junction stays in the
high-resistance (opposite magnetization direction) state on
crossing zero field (curve 2). That is, depending on the magnetic
history two stable resistance values can be realized for
$-H_s<H<H_s$.  When the field is swept from a large negative to a
large positive field value or vice versa (curves 3 and 4), the
high-resistance state is reached after crossing zero field and has
a plateau width of about 50-100\,Oe. Increasing $\theta$ results
in larger $H_s$ and also larger plateau width.

From the measured value of $\Delta R /R \simeq 3$ at 4.2\,K a spin
polarization of $P_i\simeq 77$\% is estimated. At 77\,K the TMR
effect is still above 100\%, but the rectangular shape of the
$R(H)$ curve starts to round. It still has to be clarified why the
measured TMR values are below the Julli\`{e}re value expected for an
almost fully spin-polarized ferromagnet ($P_i\simeq 100$\%).
Possible reasons are spin-flip processes arising from inelastic
transport via localized states in imperfect tunneling barriers
\cite{Klein:99a} or scattering from magnetic excitations
\cite{Zhang:97}. Such effects are not included in the Julli\`{e}re
model. Furthermore, strain effects in the manganite films may be
responsible for a reduced spin polarization close to the GB.
Unfortunately, little is known on the detailed interface and
surface properties of doped manganites. Nonetheless, the fact that
$P\approx 80-90$\% has already been observed\cite{Viret:97a,Lu:00}
indicates that a further improvement of junction and interface
quality will allow for a further increase of the TMR effect both
in GBJs and planar tunnel junctions based on doped manganites.

\begin{figure}[h]
\centering{%
\includegraphics [width=0.80\columnwidth,clip]{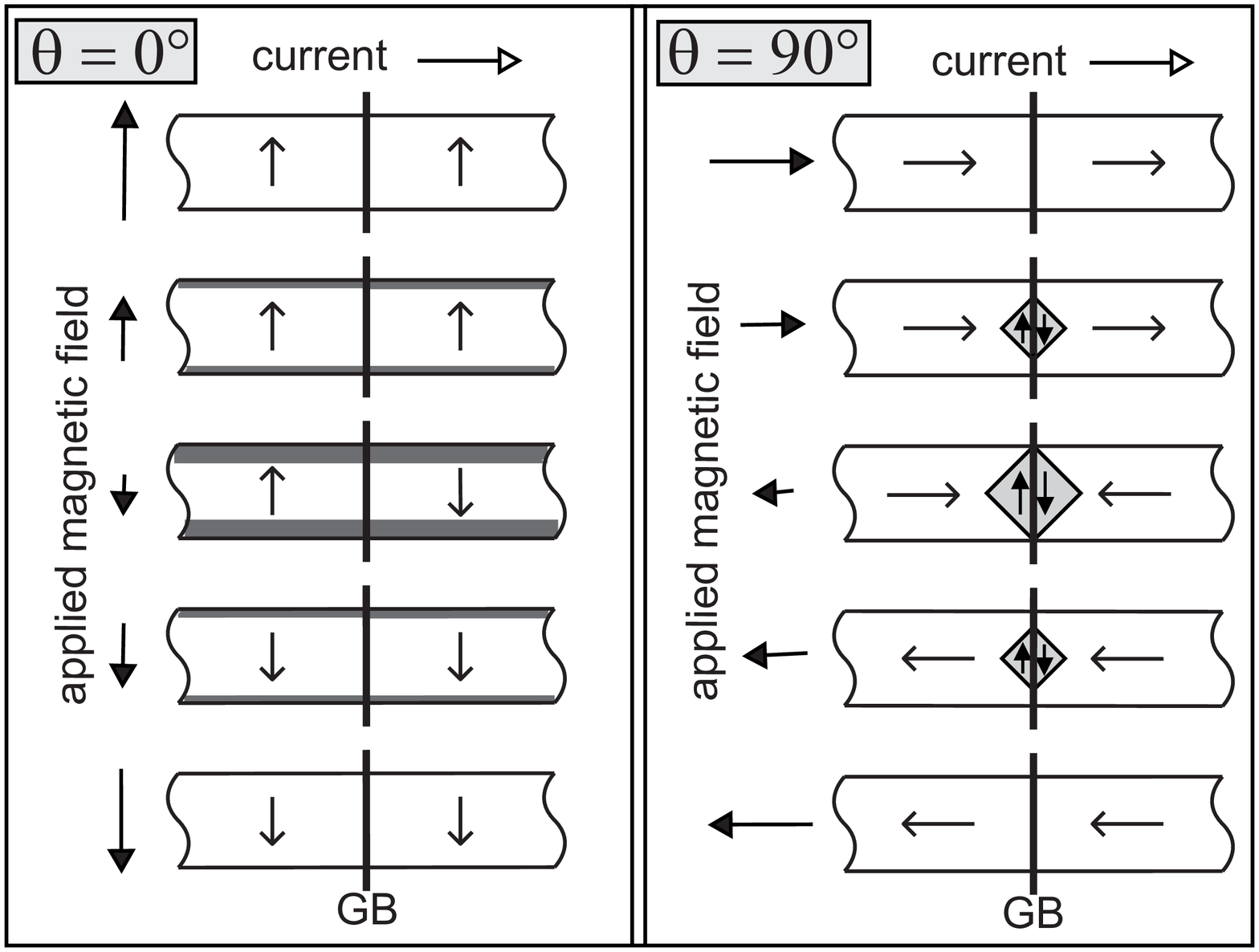}}
 \vspace*{-4mm}\\
 \caption{
Sketch of the domain structures in a manganite thin film sample
containing a single GB for applied magnetic fields of different
magnitude and direction. The shaded parts represent regions
consisting of domains with perpendicular and non-collinear
magnetization with respect to the applied field. In the left hand
panel the width of the shaded parts is overemphasized.
 }
 \label{model}
\end{figure}


We now discuss the dependence of the shape of the $R(H)$-curve on
the direction of the applied field using a simple domain model
taking into account the magnetic domain pinning at the GB
interface. A sketch of the expected domain structure is shown in
Fig.~\ref{model} for $H\perp$GB and $H\|$GB. We first note that at
the GB interface the easy axis of magnetization is parallel to the
GB. This is well known from surface studies of magnetic materials
\cite{hubert:98}. Furthermore, for $|H| \gg H_c$ the magnetization
is parallel to $H$ in both junction electrodes, that is, $R(H)$ is
low. However, on decreasing the field from $|H| \gg H_c$ a
different domain structure is expected for different field
directions. We first discuss the case $\theta=0^\circ$, where $H\|
{\rm GB}$, i.e. parallel to the easy axis at the GB. In this case,
$R(H)$ can be low or high depending on whether the magnetization
direction in the electrodes is parallel or anti-parallel.
Furthermore, on varying the field in the high-
resp.~low-resistance state, a small resistance change below 1\% is
observed due to the formation of small domains with different
magnetization direction at the edges of the sample (shaded regions
in Fig.~\ref{model}). Since these domains are small, they do not
contribute considerably to the transport behavior and
magnetoresistance of the GBJ. In a first approximation, both
junction electrodes can be considered single domain as assumed in
the Julli\`{e}re model. We note that for $H\| {\rm GB}$ the single
domains are strongly pinned by the GB interface. Thus, on
decreasing the field amplitude from $|H|\ll H_c$ a parallel
magnetization orientation and hence low $R(H)$ is preserved until
the coercivity field of opposite direction is reached. Then, at
$|H|\simeq H_c$ the magnetization direction in one electrode
switches resulting in an anti-parallel magnetization configuration
and, thus, high resistance state. This anti-parallel magnetization
configuration is stabilized in a finite field range around $H_c$
due to a reduction of the magnetic energy by reducing stray
fields. On further increasing the field also the second electrode
switches resulting again in a parallel magnetization configuration
and low resistance state.

For $\theta=90^{\circ}$, we have $H\perp {\rm GB}$, i.e. $H$ perpendicular
to the easy axis at the GB. When the field is decreased, domains form at the
GB with magnetization parallel to the GB. Since these perpendicular and
non-collinear (with respect to the applied field) oriented domains are
located at the GB, they will strongly affect the transport properties and
magnetoresistance. In this case, the electrode regions adjoining the GB can
no longer be considered as single domain. The measured triangular shaped
$R(H)$ curves can be attributed to a continuous change of the size of the
non-collinear domains. Note that for $|H| \ll H_c$ the magnetization
configuration at the GB is similar for $\theta=0^{\circ}$ and
$\theta=90^{\circ}$ resulting in similar $R(H)$ values in agreement with our
experiments.  We further note that for $\theta=90^{\circ}$ the anti-parallel
domain configuration is present along the whole GB, because the applied
field has no component parallel to the GB. Such a component destabilizes the
anti-parallel configuration and, in turn, causes a switching of the
magnetization. This explains why the $R(H)$ curves for $\theta=90^{\circ}$
form is the envelope of the $R(H)$ curves measured for $\theta <90^\circ$.

For intermediate $0^\circ <\theta<90^\circ$ the behavior can be
understood qualitatively within the same picture. An increase of
$\theta$ is found to increasing $H_s$. This is caused by the fact
that the field component parallel to the GB which drives the
switching transition decreases with increasing $\theta$. At the
same time the plateau width of the high-resistance state becomes
wider since the reduced parallel field component also drives the
second switching transition into the low-resistance state by
overcoming the energy gain in the anti-parallel configuration due
to reduction of stray fields. Fig.~\ref{angle} clearly shows that
only for $\theta>45^{\circ}$ a significant angle dependence is
observed. This is caused by the strong tendency of the domains to
align parallel to the easy axis along the GB. Certainly, the
direct imaging of the domain structure in the vicinity of the GB
would be helpful to verify our model.\cite{Vlasov:2000}  In
particular, the formation of multiple stable resistance states for
large $\theta$ that may be related to the microstructure of
individual GBs and their interface roughness has to be clarified.
Similar effects have also been observed in planar sandwich
junctions employing Co and MnFe electrodes with different aspect
ratios \cite{Gallagher:97}. So far, we have not studied in detail
the influence of different GB misorientation angles. However, we
suppose that the effect of different misorientation angles is
small compared to the effect of the relative orientation between
$H$ and the GB discussed here.

In summary, we have shown that an almost ideal, two-level resistance
switching behavior can be obtained for GBJs based on doped manganites. The
switching field, the magnitude of the TMR effect, and the shape of the
$R(H)$ curves can be varied systematically by varying the orientation of the
applied field relative to the GB barrier. The measured $R(H)$
characteristics have been interpreted in terms of a simple domain model
demonstrating the key role of a strong domain wall pinning at the grain
boundary. Our results outline possible roads for the tayloring of the
properties of magnetic tunnel junctions based on doped manganites for
applications in magnetoelectronic devices.

The authors acknowledge useful discussion with B. B\"{u}chner, R.
Klingeler, A. Marx, and S. Uhlenbruck.  This work was supported by
the Deutsche Forschungsgemeinschaft.


\begin{thebibliography}{10}

\bibitem{Julliere:75a}
M. Julli\`{e}re, Phys.~Lett.~{\bf A 54}, 225 (1975).

\bibitem{Moodera:99}
J. S. Moodera and G. Mathon, J. Magn. Magn. Mater. {\bf 200}, 248
(1999).

\bibitem{Parkin:99a}
S. S. P. Parkin, K. P. Roche, M. G. Samant, P. M. Rice, R. B.
Byers, R. E. Scheuerlein, E. J. O'Sullivan, S. L. Brown, J.
Bucchigano, D. W. Abraham, Yu Lu, M. Rooks, P. L. Trouilloud, R.
A. Wanner, and W. J. Gallagher, J.~Appl.~Phys.~{\bf 85}, 5828
(1999).

\bibitem{deGroot:83}
R. A. de Groot, F. M. Mueller, P. G. van Engen, and K. H. J.
Buschow, Phys.~Rev.~Lett.~{\bf 50}, 2024 (1983).

\bibitem{Kaemper:88}
K. P. K\"{a}mper, W. Schmitt, G. G\"{u}ntherodt, R. J. Gambin, and R.
Ruf, Phys.~Rev.~Lett.~{\bf 59}, 2788 (1988).

\bibitem{Park:98}
J.-H. Park, E. Vescovo, H.-J. Kim, C. Kwon, R. Ramesh, and T.
Venkatesan, Nature {\bf 392}, 794 (1998).

\bibitem{Lu:96}
Yu Lu, X. W. Li, G. Q. Gong, Gang Xiao, A. Gupta, P. Lecoeur, J.
Z. Sun, Y. Y. Wang, and V. P. Dravid, Phys.~Rev.~B {\bf 54}, R8357
(1996).

\bibitem{Sun:96}
J. Z. Sun, W. J. Gallagher, P. R. Duncombe, L. Krusin-Elbaum, R.
A. Altman, A. Gupta, Yu Lu, G. Q. Gong, and Gang Xiao,
Appl.~Phys.~Lett.~{\bf 69}, 3266 (1996).

\bibitem{Viret:97a}
M. Viret, M. Drouet, J. Nassar, J. P. Contour, C. Fermon, and A.
Fert, Europhys.~Lett.~{\bf 39}, 545 (1997).

\bibitem{Blamire:2000}
Moon-Ho Jo, N. D. Mathur, N. K. Todd, M. G. Blamire, Phys. Rev
{\bf B 61}, R14905 (2000).

\bibitem{Lu:00}
Yafeng~Lu,  J.~Klein, C.~H\"{o}fener, B.~Wiedenhorst, F.~Herbstritt, L.~Alff,
and R.~Gross, to be published.

\bibitem{vonHelmolt:93}
R. von Helmholt, J. Wecker, B. Holzapfel, L. Schultz, and K.
Samwer, Phys.~Rev.~Lett.~{\bf 71}, 2331 (1993).

\bibitem{Jin:94}
S. Jin, T. H. Tiefel, M. McCormack, R. A. Fastnacht, R. Ramesh,
and L. H. Chen, Science {\bf 264}, 413 (1994).

\bibitem{Mathur:97a}
N. D. Mathur, G. Burnell, S. P. Isaac, T. J. Jackson, B.-S. Teo,
J. L. MacManus-Driscoll, L. F. Cohen, J. E. Evetts, and M. G.
Blamire, Nature {\bf 387}, 266 (1997).

\bibitem{Steenbeck:97}
K. Steenbeck, T. Eick, K. Kirsch, K. O'Donnell, and E. Steinbei{\ss},
Appl.~Phys.~Lett.~{\bf 71}, 968 (1997).

\bibitem{Klein:99a}
J. Klein, C. H\"{o}fener, S. Uhlenbruck, L. Alff, B. B\"{u}chner, and
R. Gross, Europhys.~Lett.~{\bf 47}, 371 (1999).

\bibitem{Gross:99a}
R. Gross, L. Alff, B. B\"{u}chner, B. H. Freitag, C. H\"{o}fener, J.
Klein, Yafeng Lu, W. Mader, J. B. Philipp, M. S. R. Rao, P.
Reutler, S. Ritter, S. Thienhaus, S. Uhlenbruck, B. Wiedenhorst,
J.~Magn.~Magn.~Mater.~{\bf 211}, 150 (2000).

\bibitem{Hoefener:00}
C. H\"{o}fener, J. B. Philipp, J. Klein, L. Alff, A. Marx, B. B\"{u}chner, and R.
Gross, Europhys. Lett. {\bf 50}, 681 (2000).

\bibitem{Wiedenhorst:00}
B. Wiedenhorst, L. Alff, C. Recher, J. Klein, R. Gross, T. Walther, and W.
Mader, subm. for publ. (2000)

\bibitem{Gross:00a}
R.~Gross, J.~Klein, B. Wiedenhorst, C.~H\"{o}fener, U.~Schoop, J.~B.~Philipp,
M.~Schonecke, F.~Herbstritt, L.~Alff, Yafeng~Lu, A.~Marx, S. Schymon,
S.~Thienhaus, and W. Mader,  SPIE Conf. Proc. Vol. 4058 (2000)

\bibitem{Wiedenhorst:99a}
B. Wiedenhorst, C. H\"{o}fener, Y. Lu, J. Klein, L. Alff, R. Gross, B.
H. Freitag, and W. Mader, Appl.~Phys.~Lett. {\bf 74}, 3636 (1999).

\bibitem{Zhang:97}
S. Zhang, P. M. Levy, A. C. Marley, and S. S. P. Parkin,
Phys.~Rev.~Lett.~{\bf 79}, 3744 (1997).

\bibitem{hubert:98}
A. Hubert and R. Sch\"{a}fer, {\em Magnetic Domains}, Springer Verlag,
Berlin, Heidelberg, New York (1998).

\bibitem{Vlasov:2000}
V. K. Vlasko-Vlasov, Y. K. Lin, D. J. Miller, U. Welp, G. W.
Crabtree, V. I. Nikitenko, Phys. Rev. Lett. {\bf 84}, 2239 (2000).

\bibitem{Gallagher:97}
W. J. Gallagher, S. S. P. Parkin, Yu Lu, X. P. Bian, A. Marley, K.
P. Roche, R.~A. Altman, S.~A. Rishton, C. Jahnes, T. M. Shaw, and
Gang Xiao, J.~Appl.~Phys.~{\bf 81}, 3741 (1997).

\end{thebibliography}
\end{document}